**Title**: Green Scale Research Tool for Multi-Criteria and Multi-Metric Energy Analysis Performed During the Architectural Design Process

**Authors**: Holly T. Ferguson [b], Aimee. P. C. Buccellato [a], Samuel Paolucci [c], Na Yu [c], Charles F. Vardeman II [b]

**Affiliations**:

[a]  University of Notre Dame: School of Architecture 110 Bond Hall, Notre Dame, IN 46556
    E-mail: abuccellato@nd.edu
[b]  University of Notre Dame, University of Notre Dame: Center for Research Computing 111 Information
    Technology Center Notre Dame, IN 46556
    E-mail: hfergus2@nd.edu, cvardema@nd.edu
[c]  University of Notre Dame: College of Engineering 257 Fitzpatrick Hall Notre Dame, IN 46556
    E-mail: paolucci@nd.edu, nyu@nd.edu

**Corresponding Author's Details**: Aimee. P. C. Buccellato, University of Notre Dame: School of Architecture 314 Bond Hall, Notre Dame, IN 46556. Phone: (574) 631-1431, E-mail: abuccellato@nd.edu

**Name of Publication**: arXiv.org

**Abstract:**

Prevailing computational tools available to and used by architecture and engineering professionals purport to gather and present thorough and accurate perspectives of the environmental impacts associated with their contributions to the built environment. The presented research of building modeling and analysis software used by the Architecture, Engineering, Construction, and Operations (AECO) industry reveals that many of the most heavily relied-upon industry tools are isolated in functionality, utilize incomplete models and data, and are disruptive to normative design and building optimization workflows. This paper describes the current models and tools, their primary functions *and* limitations, and presents our concurrent research to develop more advanced models to assess lifetime building energy consumption alongside operating energy use.  A series of case studies describes the current state-of-the-art in tools and building energy analysis followed by the research models and novel design and analysis Tool that the Green Scale Research Group has developed in response. A fundamental goal of this effort is to increase the use and efficacy of building impact studies conducted by architects, engineers, and building owners and operators during the building design process.

**Keywords:** Building Energy Analysis Model (BEAM), Embodied Energy, LCA, Sustainability, Sustainable Data, and Building Information Modeling (BIM), Comparative Design Analysis

## 1. Introduction: Environmental Effects of Design Choices

Today, buildings, civil infrastructure, and whole cities are being designed with increasing awareness about the relative sustainability of the built environment. As of 2013, buildings accounted for 40% of domestic primary energy consumption [1] and 72% of U.S. electricity consumption, a figure that is projected to increase to 75% by 2025 [2,3]. Therefore, decisions being made today by architects, engineers, and building owners will have long-term and potentially far-reaching consequences. In order for carbon reduction strategies, such as those posed in the U. S. Energy Security Act (2007) [4] and by the Architecture 2030 organization [5], to be met, goals for limiting domestic carbon emissions by buildings need to be incorporated into *all* new building and retrofit designs conceived of today. Yet, many of the prevailing models and tools used by the Architecture, Engineering, Construction, and Operations (AECO) industry today are still reliant upon inconsistent, incomplete, and closed-source data, and in many cases, manual entry of building data for the purpose of documenting specific building metrics. Many prevailing simulation technologies require non-trivial alterations to process building models in order to conduct analyses – and most of these programs typically only perform data analysis on a single design metric at a time; for example, tools may individually be able to perform thermal analysis or structural analysis but not both at the same time. Therefore, simultaneous consideration of other metrics must often be handled by additional, distinct software tools, often requiring additional, time consuming building model alterations, thereby limiting the usefulness of these tools and their ability to positively influence the design of the built environment. Because these processes are themselves resource and time intensive, much of such analysis is conducted when a building's design is nearly complete, when many of the critical – and potentially the most resource-intensive – design decisions have already been made.

In addition to the limitations of existing analysis tools themselves, the current data deficit requires attention and novel strategies for data collection and curation methods, using frameworks that are adaptable and scalable such that they can keep pace with virtually streaming advances in building materials and construction methods. The challenges of the data deficit are further exacerbated by design time limits, the increasing complexity of building design, lack of knowledge-based design support tools, or some combination of these factors. To address these observations, the research is presented as follows:

- Identifying gaps and deficiencies in prevailing building design and analysis tools;

- Motivation for a more comprehensive and effective tool;

- Development of the Green Scale Tool (GST) for digital design and analysis;

- Testing, verification, and validation of the GST;

- Related work advancing the discoveries made in the development of the GST, including the development of new frameworks and methods for overcoming the data deficit.

## 2. Industry Standard Design Tool Deficiencies

### 2.1 Evaluating the Current Industry Standards

Building design is a complex, and essentially human endeavor, which is currently aided by digital technology (drafting and analysis tools) primarily to generate drawings and information-rich building models used to guide the construction process. Due to the increase in use of digital design and analysis tools for building design, a building's lifespan starts in computational simulation. Therefore, predictions about building performance and any consequent design decisions are currently dependent upon commercial tools and data; data that is isolated in proprietary databases, is voluminous in scale -- and growing -- and highly fragmented across disciplines.

Our interdisciplinary team began our study with a comprehensive analysis of the current state of AECO modeling and analysis platforms [6], [7], [8], [9], [10], [11] in order to discover the capabilities and limitations of these tools. There are a number of tools that endeavor to process one type of energy analysis or another [Table 1]. Revit® has individual tools that work with thermal calculations as an individual metric and recently has added a separate plug-in called *Tally* to calculate embodied energy for a given model [12]. Output for Tally extends the basic embodied energy calculation (whole or pieces of buildings) to include global warming potential, LCA [13], bills of materials, and environmental impact potentials (ozone depletion, acidification, smog, renewable versus non-renewable energy and others).

SuAT is another collaborative project that includes case studies and other influences from a variety of research groups, some of which have applications and data that are being integrated together already [14]. Some of the current calculations done within this tool are part of a Python shell and include metrics such as heat generation system types, internal heat gains (people and equipment), inlet temperatures, transmission ratios, $CO_2$ emissions, ventilation system effects, energy gain vs. losses, heating zones, and shadow-based calculations.

However, some of the most popular and often-used single-metric applications, Revit® [6], Ecotect® [7], Athena Impact Estimator® [8], Green Building Studio® [9], GaBi [15], and programs developed by the U. S. Department of Energy (DOE-2) [10] and the U. S. Department of Commerce (BEES 4.0) [11], have limited – or no – capacity to evaluate material variations against overall building impact beyond a narrow set of functions within each application. Athena, for example, is able to produce an embodied energy simulation that categorizes total energy summations into specific categories, such as transportation impacts, construction processes, and manufacturing impacts. However, it can only perform this one type of analysis, and routinely requires additional work to adjust the building model exported from Revit in order to run properly. Green Building Studio utilizes cloud-based energy analysis, including daylighting, carbon emission totals, water usage, and ventilation capabilities, as well as Energy Star [16] and LEED [17] support. Revit's thermal model plug-in called Ecotect (now obsolete) performed carbon emission analysis and thermal calculations including daylighting and shadowing calculations, albeit in a non-iterative manner, making it difficult to analyze how design changes would impact a building's overall performance.

| Existing Sustainability Tool Comparisons | | | | | | | | |
|---|---|---|---|---|---|---|---|---|
| | Green Scale Tool [29] | Tally [12] | SUaT [14] | Ecotect [7] | Athena Impact Estimator [8] | Green Building Studio [9] | DOE [10]/ BEES [11] | GaBi [15] |
| Automated and Iterative Model/Tool Interaction | x | | | | | | | |
| Zoning | x | | x | | | | | |
| Thermal Heatflux | x | | x | x | | | x | |
| Embodied Energy | x | x | | | x | | | |
| Embodied Water | x | | | | | x | | |
| LCA | x | x | | | x | | | x |
| Global Warming Potential | | x | | | | x | | x |
| Environmental Potentials | | x | | | | x | | |
| Carbon Emmissions | | | | x | | x | | x |
| Other Energy Analysis | | | | | | x | | |
| System Types | | | x | | | | | |
| Daylighting | | | | | | x | | |
| Heating/Ventilation | | | x | | | | | |
| Shadow Calculations | x | | x | x | | x | | |
| Statistical Data Analysis | | | x | | | | | |
| Climatology | x | | | | | | x | |
| Energy Star and LEED Support | | | | | | x | | |

**Table 1: Current Tool Comparison Table**

**2.2 Experiments Allowing A Measurable Perspective of Prevailing Energy Analysis Tools:**

In addition to these function-based comparisons, and in order to gain a *measurable* perspective of existing prevailing energy analysis tools, several experiments were conducted to gain an understanding of the tools' relative strengths and weaknesses for estimating key metrics of building energy use in operation and initial (embodied) energy consumption. To accomplish this, simulations were conducted with a consistent set of architectural models using the following tools: Energy Plus [18], Revit [6], Athena Impact Estimator [8], and OpenStudio [19] [Table 2]. The differences between these tool studies and model output comparisons can be seen in Table 2.

| Preliminary Industry Tool Comparisons | | | |
|---|---|---|---|
| | Embodied Energy | Operating Energy/Year | Total (100 Years) |
| Revit Energy Analysis | N/A | 1.478e9 BTU | N/A |
| OpenStudio/EnergyPlus | N/A | 2.743e9 BTU | N/A |
| Athena Impact Estimator | 2.501e9 BTU | N/A | N/A |
| GreenScale Tool | 2.369e9 BTU | 6.044e7 BTU | 8.413e9 BTU |

**Table 2: Industry Standard Tool Experiment Results**

**2.3 Observations of Current Industry Standard Tools**

The results of the experiments show large differences in outputs for the same architectural building models. This is due, in part, because settings and calculation methods vary widely between the tools. Upon further investigation, including verification by manual calculation methods, it was found that the majority of these numerical differences [Table 2] resulted from discrepancies in raw data values and the associated material properties the tools were using to tabulate the results. For example, certain tools may use significantly different density values for the same material, yielding significantly different results when aggregated over an entire building. At minimum, this suggests that the decision-support tools that architects are currently using to make critical decisions about the future of the built environment may vary widely in accuracy.

Observations of the outputs of these tools (i.e. aggregated energy use totals, for example) demonstrate that the building simulation programs used by architects today lack uniform data and processes to facilitate the consistent and reliable evaluation of both customary and novel building practices. The consequences of consistency and reliability impact the environment from the commencement of the design process, to the selection of materials, the methods of their assembly, and the long term implications of a design on the environment – alongside lifetime building operating energy use, the most common measurement of building energy performance.

## 2.4 The State of Data-enabled Building Design

In addition to tools, existing gaps in data increase uncertainty by the user and meanwhile lead to a widening gap between simulated and achieved building performance. However, this is not necessarily the case in the data-rich disciplines of science or even in more closely-allied engineering disciplines. In fact, when compared to other disciplines, like automotive and aerospace design, or closely allied industries like civil engineering, architecture underutilizes current computational and Big Data analytics technologies. If such methodologies were harnessed in the building professions, by the architects and engineers who design and create the built environment, we could achieve a much greater understanding – and practical ability to effect – the way that energy and resources associated with the built environment are consumed or conserved. In addition to GS research of prevailing methods and tools to measure the impact of buildings on the built environment, the work has advanced understanding of the data currently used in data-driven design analysis, inspiring subsequent phases of research: the development of better analysis tools and more informed approaches for achieving robust decision-support for sustainable building design.

## 3. Motivations for a More Comprehensive Tool

The GS Method and Tool – and the subsequent case studies conducted with it – advance conclusions drawn from allied industry case studies: that single-metric analysis fails to capture the full energy impact of a building over its lifespan. In fact studies have shown, the material embodied energy involved in the construction of purported "low-energy buildings" can account for up to 46% of total life cycle energy use, demonstrating that a reduction of energy use in building operation does not always result in a reduction of overall (net) life-time energy use [20]. Furthermore, since up to 90% of the total life-cycle cost of a product is determined in the early stages of design [21], critical and timely access by AECO professionals to accurate and reliable embodied energy data – and tools that can meaningfully integrate this data into a design process that also takes operating energy into consideration – means greater potential for that information to positively influence decisions made *during* the design process, when those decisions can have the greatest overall impact on building net energy use.

Our observation of, prevailing analysis tools is that most lack the capability to accurately or uniformly quantify energy use tied to material extraction (production, transportation, and assembly) *and* energy use in operation, along with fundamental site and climate-related factors. Industry-leading digital modeling and whole building carbon analysis software include Revit® [6], Ecotect® [7], Athena Impact Estimator® [8], Green Building Studio® [9], and programs developed by the U. S. Department of Energy (DOE-2) [10] and the U. S. Department of Commerce (BEES 4.0) [11]. These software packages are presently not capable of accurately evaluating whole building impact, particularly those impacts related to material choice at the level of an individual component or unique assembly, but only according to basic palettes of predetermined assemblies. This rigidity limits their use in guiding informed choices about the adoption of emerging materials and technologies. To attain a fuller perspective about how different energy metrics (for example, thermal flux, versus embodied energy, versus cost, etc.) might effect a building's overall performance, modern architectural design and analysis tools must be built to achieve greater parametric analysis functionality than is currently allowed by the most widely used single-metric applications. This is the motivation and framework for the development of the Green Scale Digital Design and Analysis Tool (GST), described in the following sections.

## 4. The Green Scale Tool (GST)

### 4.1 Green Scale Tool: A New Digital Design and Analysis Tool to Achieve Simultaneous Analysis of Embodied Energy and Annual Thermal Loads

The GST architecture is presented in Figure 3. As the figure describes, the GST is designed for compatibility with Autodesk Revit®. Currently, Revit uses the gbXML file standard [22] to describe and export any model from its 3D rendering space. This process is automated through the GST plug-in followed by the main algorithm call to run the specified energy modules [Figure 3].

First, the gbXML file for the given architectural model is parsed into a data structure that can be queried for the various material data and thermal property values. Following the data structure creation, subsets of modules are activated depending upon the user settings. For example, the GS Thermal Model can access Energy Plus climatology data along with the calculation routines and the GST Embodied Energy Model can access the requisite external Materials Database. After the calculations are completed, user log files are generated containing the relevant data for each design metric, reformatted for presentation to the user. Finally, the GST application interface is continuously updated with the new results of various comparative iterations. Intelligent decision support [23] mechanisms could be employed at this point in the process to give further recommendations about material choices. This iterative feedback loop continues as needed by the user.

Our effort contends that in order to make the best, most-informed building design choices, energy, thermal impacts, cost, and other factors should be considered *simultaneously* in a design process – and the most effective analysis tools should promote this. The modular software architecture embodied in the GST enables multiple simultaneous calculations and comparisons. The GST currently conducts thermal analysis by calculating heat flux and it also concurrently calculates material (embodied) energy impacts through an embodied energy model. When the geometry data is loaded into the tool from a standardized gbXML format, it gives the building a computational mapping [Figure 4] from which to read data for running the energy simulations. To achieve this level of analysis the Tool divides the building into individual spaces, surfaces that make up each space, and finally into single materials of each surface assembly [Figure 4].

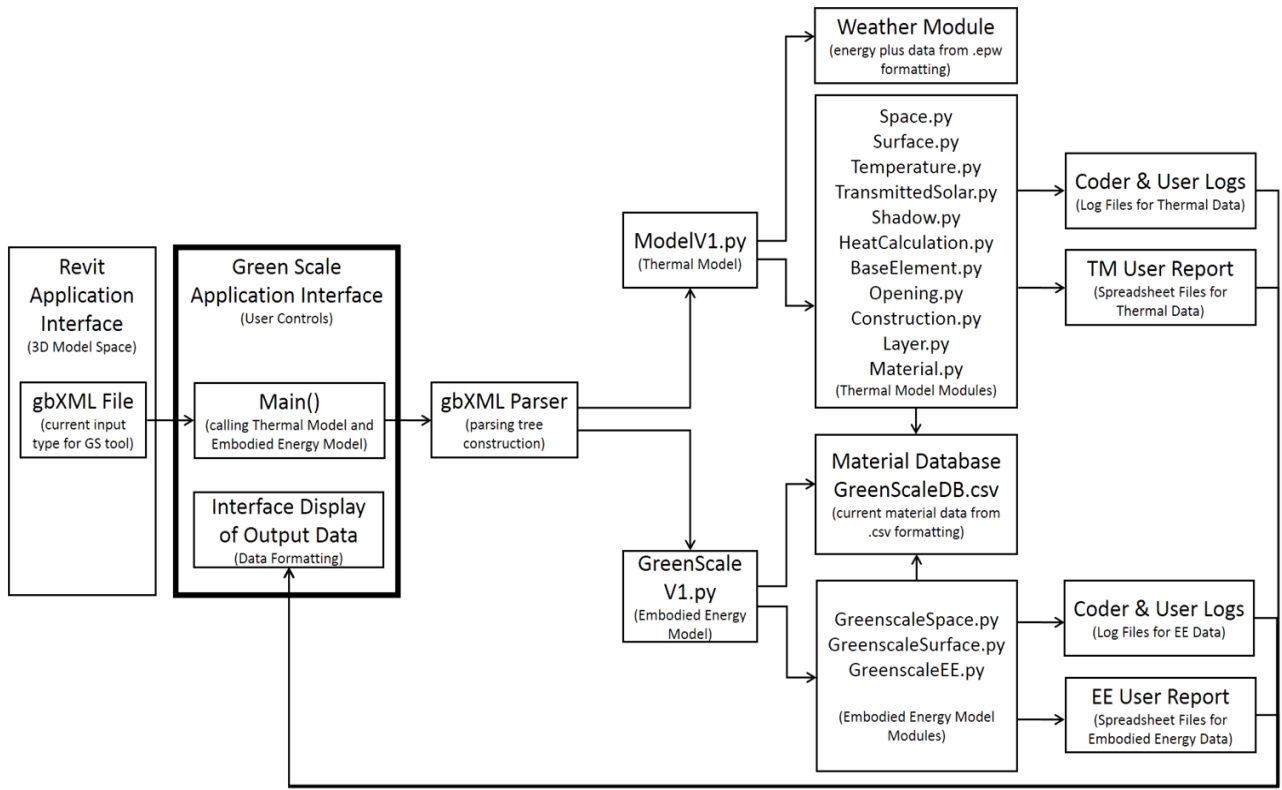

**Figure 3: Green Scale Tool Computational Architecture**

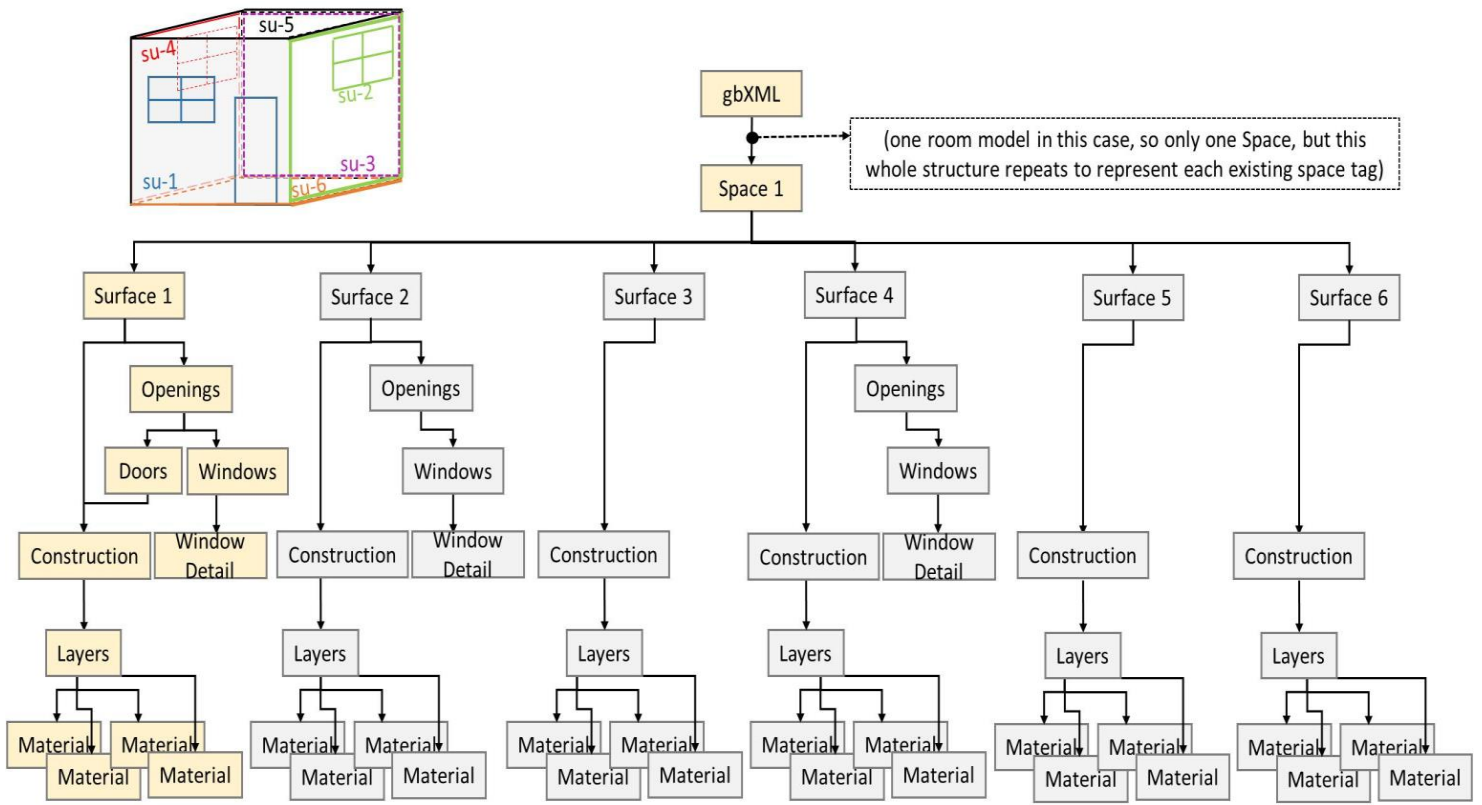

**Figure 4: Architectural Model breakdown by the GST for Computational Processing**

## 4.2 Green Scale Methods, Graphical User Interface (GUI), and Overall Application Architecture

The Tool is a modular Python framework [Figure 3] that facilitates data interchange with industry standard tools, currently Revit, and is specifically designed to connect architectural models with various performance metrics. The Tool's initial function modules were built for on-the-fly calculations of embodied energy and thermal performance characteristics that can be run from a Revit plug-in [See Appendix A]. For full details about the implementation of the GS Graphical User Interface (GUI) and overall application architecture, please see Appendix A.

### 4.2.1 Experimentation Method and Motivations

The modules implemented within the Tool will eventually be capable of connecting to any external database that conforms to a predefined web interface standard and contains material properties and parameters required by the models. This design feature allows very flexible configurations of data sources. These sources can range from local database(s) to cloud-based or crowd-sourced datasets. The implementation presented in this paper utilizes a centralized database that facilitates controlled evaluation and testing of the tool from reliable data.

The standardized gbXML [22] format is one of the ways architectural modeling data can be transformed into a computer readable template that can be utilized by the GST. Revit readily provides this as an export option that can be automated for use in the tool; additional file formats are intended to be linked to the tool in future versions. This means that the GST will be able to work with a number of architectural model source outputs in combination with the independent materials database. The extended architecture would allow execution of this and other design evaluation models on distributed computing environments as design complexity increases.

### 4.2.2 Green Scale Thermal Model Description

The GST uses a simplified heat flux methodology that utilizes a lumped capacitance model [24] for computational efficiency to provide faster feedback to the users while still maintaining reliability of the results. Although the model requires only a few parameters, experienced users can utilize data-rich options if desired, such as a convection coefficients, time step granularity, and terrain specifications. Since the Python scripting language is interoperable with other platforms and languages, it was the chosen language for the GST Thermal Model module.

Derived from the Building Energy Analysis Model (BEAM) [25] with implementation written in MatLab, the GS Thermal Model offers rapid analytical feedback by using accurate methods for predicting energy usage, using two different time steps for climatology data: every hour for every day and every hour for every other day. The latter of the two time step protocols is able to save 50% of the algorithm processing time. Employing the Euler explicit method, the time dependent terms in the code can be integrated using the previous time step's temperature information throughout the calculation [25]. A larger time step than every hour for every other day does decrease the run time but increases the fluctuations in heat flux depending upon the location by as much as 10% compared to the every hour for every day granularity [25]. For example, where a building in Emmonak, AK maintains the same heat flux under a larger time step, buildings in Washington DC and Houston, TX are too sensitive to the same time step because in these cases too much weather information is omitted [25]. Therefore, the best compromise to maintain accurate calculations for any location while getting the fastest possible processing time resides with the every hour for every other day time step protocol.

For the Thermal Model module of the GST, unit tests were implemented as a secondary verification against the BEAM model. Data structured in a gbXML format gives the geometry and associated material information of a building that is needed for space heating and cooling load calculations. Each building is calculated by dividing it into components (i.e. walls, floors, windows, and doors) with different boundary conditions. For each of these building components, the three modes of heat transfer -- conduction, convection, and radiation -- are used to calculate the heat balance. Adding internal heat sources is one of several future modules [25] to be implemented in the Tool. Monthly and annual energy demands are output by the GST, the accuracy of which is discussed below in the verification and validation section. Additional modules that can be added to the BEAM/GST could include a mass balance analysis (humidity), ground temperature, occupancy, electronic equipment, structural, HVAC, etc.

### 4.2.3 Green Scale Embodied Energy Model Description

The Embodied Energy (EE) is calculated by using the geometries from the standard gbXML file in conjunction with material property values queried from the GS Materials Database. The calculation process is achieved by dividing the model into spaces, then surface assemblies, then finally into each individual material [Figure 4]. These individual materials are then re-aggregated into primary building assemblies that support the user in their design choices. Embodied Energy is calculated for each material as the product of material EE, its volume, and density. Finally, an EE summation for each assembly is tracked and added to a current total until a building total EE is found; additionally, the Embodied Water (EW) can be found based on that total. The values for EE and EW can be output as building totals, assembly aggregations, by individual surface within the building spaces, or by individual materials used. The current tool requires manual intervention to supplement materials between the gbXML file and GS Materials Database, but future versions of the tool will allow automatic data integration [26].

## 5. Green Scale Tool Verification and Validation

Since creation of the GST was completed, the research proceeded to test if the Tool was more reliable, effective, and usable than prevailing analysis tools used by the greater architectural community. Several comparisons and case studies were performed to prove accuracy of the tool as well as to measure data outputs against the aforementioned leading industry tools (EnergyPlus, Athena, etc.).

### 5.1 Green Scale Thermal Model Verification and Validation

The verification of the BEAM algorithm, in both the Matlab and Python GST, was based on three basic building structures: single, two, and four room models [Figure 5]. Models have been verified against several locations (for varying latitude and longitudes) around the United States (Washington DC, Houston, TX, and Emmonak, AK). Weather data is used directly from the EnergyPlus Typical Meteorological Year 3 (TMY3) files [18]. The Matlab verification and the GS verification are described below.

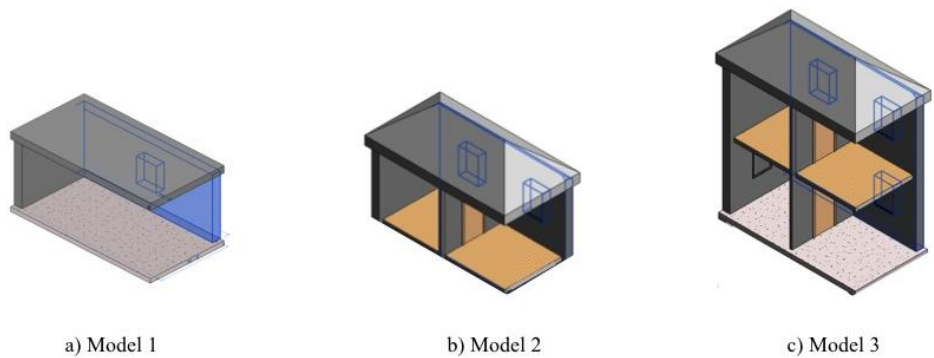

a) Model 1           b) Model 2           c) Model 3

**Figure 5: (a) Model 1 as the Single Room Model, (b) Model 2 as the Two Room Model, and (c) Model 3 as the Four Room Model**

### 5.1.1 BEAM Technology Validation

The original implementation of the BEAM algorithm, written in Matlab, was established in comparison with EnergyPlus [18], the industry standard. This standard has well-documented advantages over other simulation tools used in the architecture industry [27]. EnergyPlus bases its calculations on popular features and capabilities of Building Loads Analysis and System Thermodynamics (BLAST) [28] and DOE-2, allowing an accurate and integrated simulation/analysis.

In the GST development experiments, BEAM is used to evaluate three different geometric models (single room, two room, and four rooms on two levels) each of which are tested in three different geographic locations around the United States. The results of BEAM are validated by comparing outputs with the results from Energy Plus for each of the respective geometric models. The settings for each Energy Plus simulation are as similar as possible to the methods used in BEAM. Results for the three models are presented [25] [Figure 6]; these models were created in Revit Architecture and displayed in SketchUp. All show correct correlation between the Energy Plus simulations and those from BEAM for the respective models; however, the runtime for the BEAM model (once translated into the GST) is faster than EnergyPlus and can also be more easily integrated with additional popular BIM design tools (the single room model is 52.0%[1] faster, the two room model is 33.02% faster, and the four room model is 43.05% faster than EnergyPlus).

Figure 6 shows the effects of a correction multiplier α for the monthly thermal loads in the three test models and locations. The correction multiplier is the ratio of effective thermal capacitance over apparent thermal capacitance. Generally, this means that α will be between 0.32 and 0.45 for insulated buildings and approximately 0.25 for uninsulated buildings. The single room model, reducing α tends to increase the monthly thermal loads regardless of the location. The effects are similar for the two room and four room models. Heating and cooling loads reflect more sensitive changes from the correction multiplier when outdoor and indoor temperatures are in close range to each other or when there are large fluctuations in temperature between day and night [25]. An example of this effect can be seen during May and September in Washington, DC as well as April and October in Houston, TX. It should also be noted that when the temperatures are fluctuating near the set room temperatures, both heating and cooling are utilized alternately making the thermal capacitance changes more evident. Additionally, the time period causing the biggest temperature differences resides at the change between day and night when thermal capacitance is the most influential [25] [Figure 6].

---

[1] Model run times are subject to certain settings in EnergyPlus, thus these run times exist relative to the settings chosen.

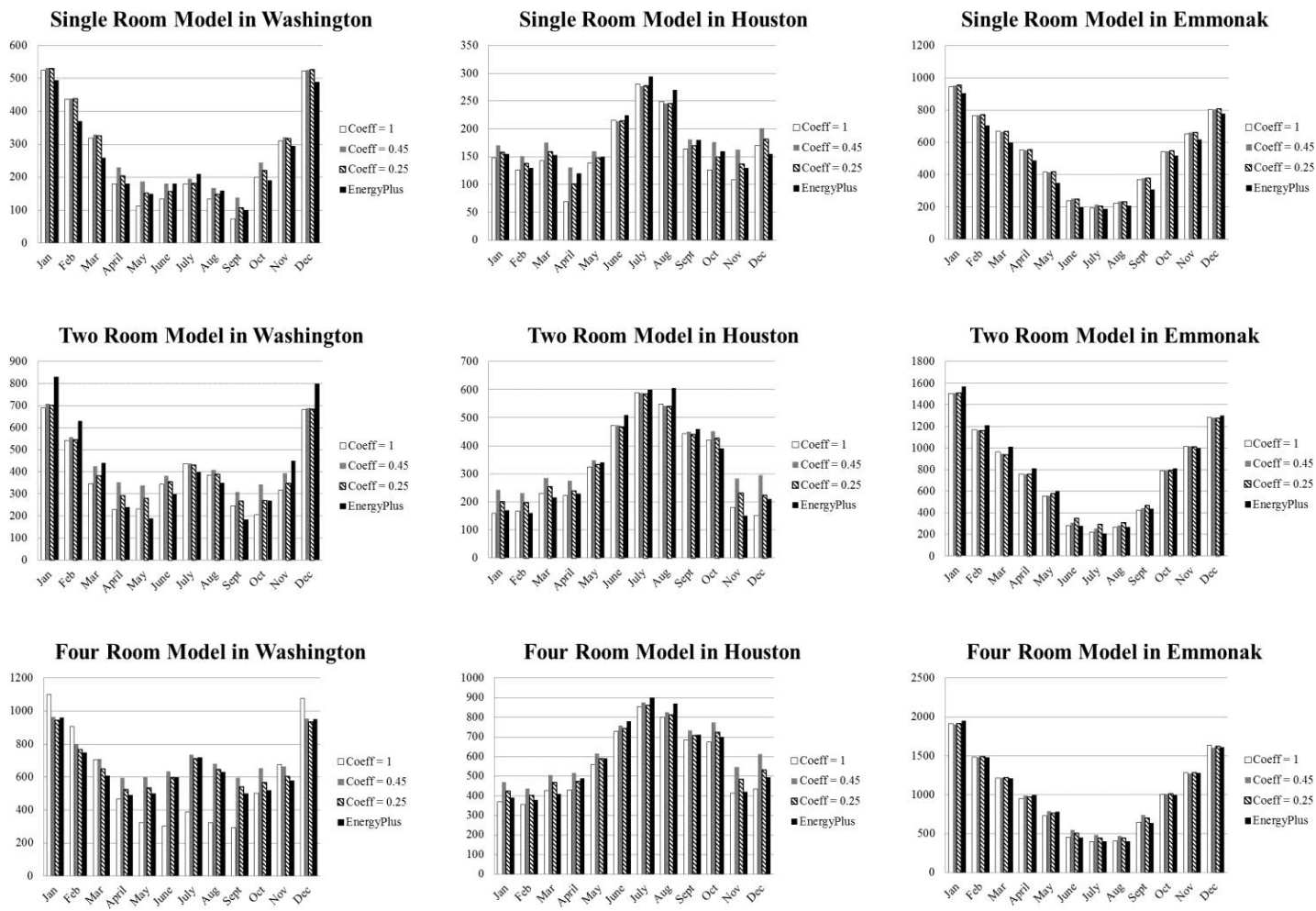

**Figure 6: BEAM Monthly Thermal Loads**

### 5.1.2 Validation: The Green Scale Thermal Model

For validation of the GS Thermal Model, the time-step protocol used in BEAM and the Energy Plus weather data was set at every hour of each day. The GST was set without shadow components and at three convection coefficients of 1, 0.45, and 0.25, each for the same three locations around the United States (Washington DC, Houston, TX, and Emmonak, AK) [See Appendix B]. A terrain type of Flat or Open Countryside was tested for these simulations. The Thermal Model output currently reports data in kWh used per month (heat flux) and for the total year [Figure 7].

As seen in Figure 7 and in all charts presented in Appendix B, the GST output data correlates exactly with outputs from the Matlab BEAM model for the tested case studies. This demonstrates that the GST is calculating using the same, validated algorithm produced in Matlab for BEAM. The identical output is shown for the same three models tested against EnergyPlus.

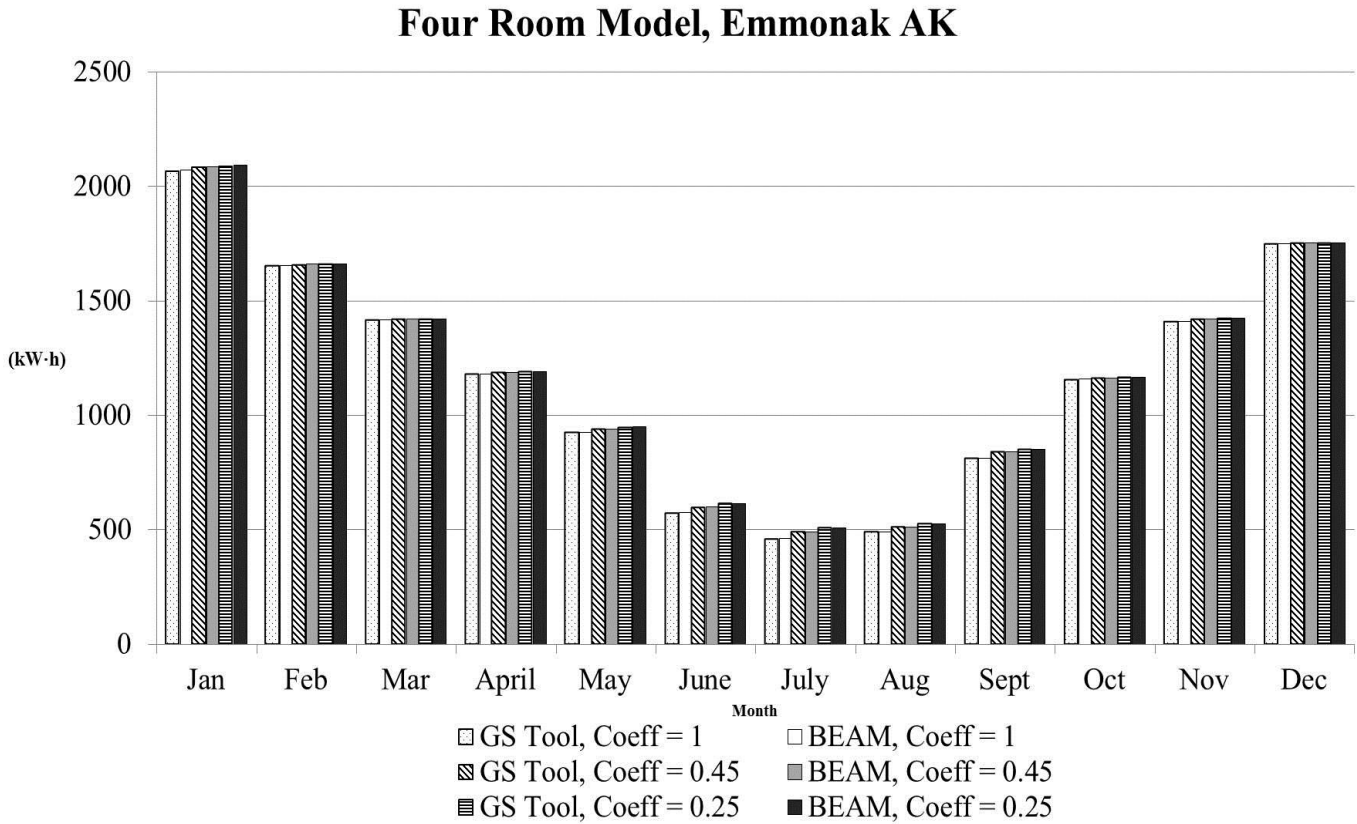

**Figure 7: Four Room Model heat flux comparison between the GST and BEAM for Emmonak, AK, at three convection coefficient settings**

### 5.1.3 Discussion: Green Scale Thermal Model

The algorithm used for the GS Thermal Model is a Python translation of the BEAM Matlab algorithm. While the overall algorithm does follow the same programmatic steps and correlates well, there are still differences between the GS and BEAM output for yearly totals falling between $2.46 \times 10^{-5}$% and 0.03% [Figure 7]. This is not significant when machine precision differences, differences in the two language structures, and rounding within pre-existing library functions are considered. This fluctuation would be seen even if the library functions were not rounding until six places after the decimal. In other words, this means that the GST is performing the same as BEAM for these three models, in the three chosen locations, and for the same range of convention coefficients [See Appendix B].

These case studies reveal the accuracy of the GST consistency with Energy Plus output, and calculations. The GST gives a body of data regarding the predicted thermal capabilities of architectural building models; this data can be evaluated simultaneously with the Embodied Energy (EE) Model. Effects of thermal gains and losses throughout the entire calendar year offer one important perspective about materials that should be considered when making design choices. Now these effects can be tabulated and measured along with other metrics of the GST at each step of the design process.

### 5.2 Verification and Validation: The Green Scale Embodied Energy Module

Validation of the EE algorithm in the Python GST was based on two basic building structures: the single room model used in the thermal validation and the Robert L. Miller Veterans' Center (RLMVC)[2] model consisting of 182 gbXML surfaces (including concrete foundation walls) which bound 30 different spaces constructed at ground level [29]. The RLMVC case study provides an example of an existing real-world construction that is able to maintain calculation integrity. The external GS Materials Database, which both of these models use, is accessed through a web service with its own data preservation verification methods.

The validation process included running building design models constructed in Revit through the Tool and creating manual spreadsheets of the calculations for comparison. The spreadsheets systematically totaled the EE for several models (surface-by-surface and assembly-by-assembly) by following the same steps that the corresponding Python code uses. The sheets were organized by dividing the model into surface aggregations and then accounting for each individual material in that surface. Embodied Energy is calculated within the spreadsheets for each material by manually applying the same formula described earlier. Once the Tool total output matched the manual calculations, types of surface assemblies were aggregated within the spreadsheet and compared with the assembly aggregations tabulated by the Tool. Finally, the calculations between the GST and the GS manually aggregated spreadsheets were compared against the output of a related industry standard tool used for gathering EE data, Athena [8]. The results for the models correlate well and can be seen in Figure 8.

---

[2] The RLMVC is a 5000SF transitional housing facility for homeless US Military Veterans located in South Bend, Indiana and designed by author, Aimee Buccellato.

### 5.2.1 Green Scale Embodied Energy Model Validation

To verify the material property values in these two case studies (single room and RLMVC), various external databases were referenced with the appropriate value conversions made as necessary within the Tool. All materials in these spreadsheets were sourced and recorded using the various sources for material property information; this gave a method of validation to the EE case study materials used in the computerized simulations. The property values of the materials used in the models were thoroughly researched by the GS Team for accuracy and consistency. This process involved resolving discrepancies for variations; variations that routinely occur between several reliable sources of material data properties including ASHRAE [30], IFC [31], BIM [32], and others.

The current EE model relies on the validation of the material property values existing in the given gbXML file as well as the centralized GS Materials Database. Conflicting data sources, units of measurement, naming schemas, curation techniques, and formats make decisions about data correctness a challenging process. Unit Tests are used to verify the EE model code and compare to manual calculations of the single room architectural model. The manually calculated single room and RLMVC architectural models were completed using a similar process as the code itself, including the material property values from the gbXML and EE values from the database. Data validation has been completed for the materials in the GS Database with sourcing tags for each value; additional materials are in the process of being added to the database and sourced as they are used in architectural models generated as part of this research.

Figure 8 presents the values of the validation process for the single room model as well as the RLMVC case study. Between using gbXML file data from Revit as well as material properties from the independent GS Materials Database, the results of the EE calculations are correlating for both of the tested models for the manual spreadsheet calculations versus GST totals.

As an extra validation, total EE values from Athena for the respective models are compared to the totals from both the manual spreadsheets and the GST. This general correlation shows that the GST is calculating to the standards of other existing tools for the tested models [Figure 8]. The discrepancies in this final comparison do not necessarily exist due to issues with the GS Model, but instead can be attributed to a variety of factors discussed in the next section. The GST comparison against a known industry method of tabulating EE (both manually and computationally) provides evidence of the capability of the GST to provide useful data aggregation related to EE (and embodied water) throughout the whole architectural design process.

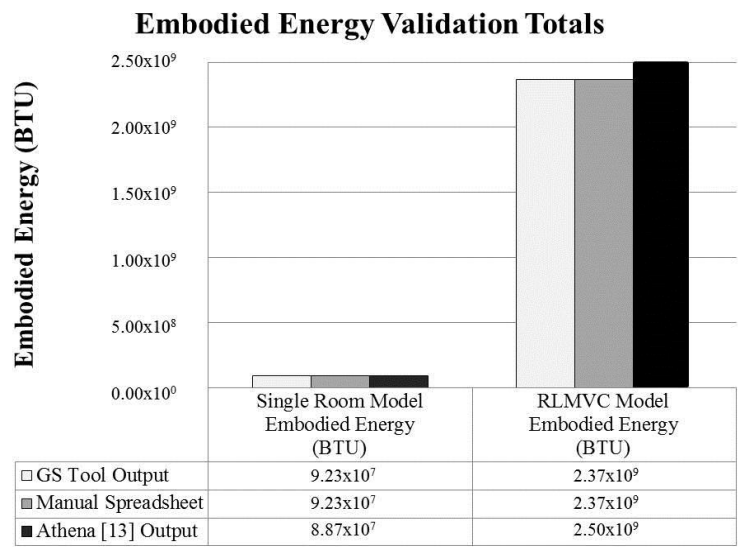

**Figure 8: GST Validation Comparisons between Manual Spreadsheets and GST Output**

### 5.2.2 Green Scale Embodied Energy Model

The materials in the single room model have a total EE of $9.23 \times 10^7$ Btu calculated manually in the spreadsheet and also $9.23 \times 10^7$ Btu calculated by the GST [Figure 8]. Similarly, the materials in the real-world RLMVC model have a total EE of $2.37 \times 10^9$ calculated manually in the spreadsheet and $2.37 \times 10^9$ calculated by the GST [Figure 8]. As the models become more complex, there is a larger error margin due to several factors: one of the sources of error seen is the result of translation between two different calculating protocols – manual spreadsheet formulas and Python functions. Each language uses a different methodology for handling data types, value precision, and the algorithms used from libraries as built-in functions. Each tool uses completely separate sets of libraries; for example, this means using an operation in one tool such as "multiply" may have a different protocol of precision than the equivalent operation in the other tool. Even if no other sources of error existed, this minor difference would accumulate over the course of aggregating EE for a whole building, causing a variation like the one seen for the RLMVC. However, there are other potential sources of value differences. While both the GST and the manual spreadsheet method use a certain level of precision, manual rounding as each of the values are entered into the spreadsheet also occurs. This human error – or assumption that values are precise enough as entered – can also cause a noticeable fluctuation in the final result. Additionally, there was also a precision inequality depending upon how the unit conversions were handled. This happened in certain cases along with the values further being rounded by the aforementioned precision limits of the spreadsheet. These number conversions did not impact a single operation by a noticeable amount, but once propagated across a whole building, these can become noticeable. This would not be isolated to a research case, but could certainly be seen in manual EE analysis in practice, with the potential to propagate either overly generous – or conservative – anticipated EE for a structure.

A double verification was preformed (i.e.) a second iteration of the RLMVC spreadsheet to compare the differences between rounded values (up to the sixth place value). This did prove human rounding and unit conversion as sources of minor value discrepancies between the spreadsheets and the GST; this means that the differences are not a fault in the GST calculation procedure. Discrepancies such as those seen in Figure 8 are understandable due to their sources; therefore, the GST is still an accurate means of tabulating the EE for a building.

Athena gives $8.87 \times 10^7$ (1.04% difference between GS and Athena) Btu for the comparable single room model and $2.64 \times 10^6$ (1.05% difference between GS and Athena) for the RLMVC model; comparable because the simulation for the single room model created in Revit was the same model run in Athena with only the material alterations and edits Athena requires. This accounts for a percentage of the error between Athena and the GST output for both validated models. For this validation, it is necessary to alter certain input models (Revit) to fit the format of the Athena program; this is one specific example of why more efficient energy analysis tools are necessary in order to effect change *during* a normative design process. Nevertheless, this also requires a certain amount of alteration to the naming and designation of certain materials used in Revit, yielding different thermal and EE properties found for each change. Also, since neither the source code nor database(s) for Athena are available, it is difficult to directly compare any calculations made therein to the GST, in terms of process, sources, and/or calculation precision. The only breakdown in data aggregation from Athena is for the total EE; this total is divided into manufacturing EE, product transportation, construction transportation, and construction installation. However, the reliability of these figures is unclear unless the algorithmic processes and data sources become available.

Undoubtedly, the largest source of discrepancy between GST and Athena occurs because material property values are queried from different databases. Without a direct comparison between the two for each material, it is unclear if values used in calculations are at all comparable. It is also not openly known where Athena – and most other tools – retrieves its material information, including source reliability, differences in manufacturing, and unit conversion precision. Another reason differences in outputs are expected stems from special case conditions where calculation processes cannot be directly compared to Athena's, since the processes are not public. For example, certain surface types such as parapets, foundation, partitions, and overhangs are not identified in Athena in the same manner as space-bounding surfaces are identified in the standard gbXML output from Revit. Since each of these conditions existed in the experimental cases, uniform approaches for handling them were needed in order to arrive at a logical construction type for each, and material data for calculations for embodied energy. This means that the GST can be compared to manually aggregated assembly totals, but without an open source method of Athena's algorithm, the comparison between these two tools is challenging.

Regardless of variations seen in EE due to the reasons explained above, all three sources (GST, manually aggregated spreadsheets, and Athena) are successful and conclusive for verifying the usefulness of the EE calculations. When the GST uses this module in an automated, iterative method in conjunction with a Revit model, the continuous feedback will provide invaluable, multi-criteria decision support from the beginning of a design project, enabling design choices that can be made simultaneously with reference to material energy costs and anticipated thermal impacts.

**5.3 Architectural Information Model Coherency**

Validation of correctness and completeness of exchange and extraction of architectural information with respect to the schema used between the Tool and Revit was also studied. Motivation for this analysis was the noted discrepancies between hand-based analyses and the tool analysis for previous case studies. One prominent source of this discrepancy was narrowed down to the differences in interpretation of architectural objects in studies conducted by hand versus the representation encoded in an information schema. For the purposes of understanding this variation in results, validation was completed on four Revit models of increasing schema complexity (Single, Two, Four Room, and RLMVC Revit Models). Accuracy, readability, and complete formatting of the exported gbXML files was verified, as well as the corresponding unit tests when the architectural models were run through the Python thermal models. It was determined that additional procedural rules were needed to adjust the Tools interpretation of a particular architectural situation represented in schema versus a practitioner's implicit knowledge of that situation. More details on these rules and their implementation in the Tool can be found in Ferguson et al. as well as a more transferable approach using a formal rules engine. It is envisioned that the situational and conceptual knowledge can be formalized using an ontology-based approach that would facilitate the construction and sharing of such knowledge bases. An additional series of test architectural models were constructed to validate other types and/or extensions of verified geometries that contain unique geometries or structural situations; certain problematic architectural models found are still being analyzed. As these circumstances are discovered, additional rules are added to the tool to resolve the problematic geometries and schema.

## 6. Current and Future Research

### 6.1 Future of the Data-supported design, the Green Scale Tool and Related Components

It has been observed by Dixit et al. that existing data sources for total life cycle energy analysis, like embodied energy, for example, vary widely in terms of quality and do not account for geospatial dependencies of measurements within the data [33]. In addition, this data is broadly heterogeneous, existing in a wide variety of data formats or within siloed, proprietary databases – much of it restricted material data unable to be accessed publically. However, the incorporation of this information would vastly improve the results and precision of the GST output, and increase the efficacy of the Tool – and tools like it – to positively influence design decisions made using building information and data.

Another common problem with data in use today manifests itself as missing or incomplete in content and attributes. Even if the data is subject to an effective organizational schema, there is an issue of handling missing pieces or disjointed sets of data. Ontologies and patterns can solve this problem and research is underway by the GS Group to study these solutions; one such solution is called Material Transformation Pattern [34] that can be used to map the energy costs involved in material state transitions that occur during material manufacturing, shipping, and the building process itself. These types of ontological patterns also allow connections to be made between themselves and other related patterns such that they can work together. This means that even more types of data can be evaluated, thus generating a much more comprehensive picture of total building impact.

### 6.2 Independent Green Scale Material Database

The authors believe that the integration of mechanisms that facilitate data contribution within a tool that is in close proximity to architectural design activity can address aspects of these issues. Information manually entered into the GST can be contributed, with proper provenance and attribution, to a centralized repository available to all data consumers. Contributed data can be peer reviewed for accuracy and quality providing some context for other data consumers. The data store could be later extended, utilizing World Wide Web Consortium (W3C) semantic web standards [35] to share the contextualized data, metadata, and provenance information that can be consumed by other agents or tools. Several groups are already exploring these technologies to address various aspects of the sustainability problem [36, 37, 38]. Now, additional capabilities can be introduced to improve tool functionality such as rule engine integration [39], crowd sourcing to gather architectural knowledge, and web crawling techniques to acquire new material data. By unlocking the data and metadata, building industry professionals can use tools, like the GST, to positively impact how the built environment is created.

### 6.3 Green Scale Model Additions

The authors propose – and are currently pursuing – the development of additional modules that can be used in conjunction with the GS Thermal and EE Modules to generate a larger range of comparative metrics between different designs. Material costs, for example, can be introduced by using features from the existing EE Module. The architecture developed for the GST will enable additional modules to be added to the Tool as the research progresses. Of course, any additional metrics will also be dependent on data sources, in many cases, specific manufacturer listings in different regions. However, this data could also utilize novel ontology structures and decision-support features under development by the Green Scale Group. Additionally, certain Revit object features have not yet been included in the existing GST Modules, but can be included in the development of additional features in the future, including:

- Additional CAD object identification
- Additional structural elements/features which will be contributing factors to the various computer model EE calculations
- Mass balance analysis (humidity)
- Occupancy and scheduling data
- Electronic equipment
- HVAC system elements,
- and other customized modules

By extending the GS Materials Database with ontological assistance, and by simultaneously increasing the extensibility of the GST by incorporating additional energy model types (such as those listed above) that give a wider range of calculation metrics, comparative analysis and decision-support is greatly advanced for architectural design decision-making. As a result of these current and future research endeavors, more informed design decisions can be made, and the broader impact of those decisions on the environment can be anticipated, mitigated, or avoided altogether. The authors of this work realize that the complexity of the tool will grow with the amount of data and the number of design objectives that are to be taken into account. In such cases more advanced methods for multi-objective optimization, such as interactive evolutionary computing methods [40] can be used.

## 7. Conclusion

Current computational tools used within the architecture and engineering professions aim to gather a thorough and accurate perspective regarding the impacts caused by the built environment. However, as this research presents, in order to gain this fuller perspective – and fully understand the both the overall and granular impacts of different building design decisions, modern architectural design and analysis tools must be built to have more concurrent and comparative functionality between design metrics. Currently, the most widely used applications often only utilize a single analysis metric. The Green Scale Tool combines thermal, embodied energy, and lifetime cost analysis within a single interactive interface, such that a variety of metrics can be compared and considered simultaneously. This approach gives designers more immediate access to a broader range of data and knowledge from which to make more informed design choices, from the very beginning of the design process (in simulation) and throughout the design process, with potentially more far reaching results for environment.

The thermal model used in the GS tool demonstrates correct correlation compared to the BEAM thermal model for all models, locations, and other parameters tested. The EE energy model also maintains accuracy for calculations down to each material layer for both simple models and existing buildings used as case studies. Using a modular structure, these metrics can easily be combined with additional ones, gaining an even fuller perspective on building material choices from the beginning of the design process, allowing architects to have a fuller perspective of the impacts of their design choices.

The authors believe that if more accurate and reliable building data can be sufficiently structured, and opened to enhance novel tools and infrastructures for data and knowledge acquisition, it will enable building endeavors to achieve a better built environment. This approach also fosters the discovery and sharing of knowledge between communities that can further increase data accessibility and improve positive environmental impacts. Use of models and languages capable of capturing collective knowledge will enable sharing within the larger professional community and simultaneously encourage community member contributions via crowd-sourcing leading to a higher probability of achieving the collective goal of a more sustainable built world. The GS research group envisions a future that develops this community involvement to continually improve data reliability in order to achieve truly sustainable and energy efficient buildings in a potentially unprecedented fashion.

## 8. Acknowledgements

The Green Scale Research Project has been made possible by the University of Notre Dame Office of the Vice President of Research Faculty Research Support Program 2012 Regular Grant and the University of Notre Dame Center for Sustainable Energy Sustainable Energy Initiative Program and undergraduate research support by the Slatt Foundation. The GS Group is also grateful to the contributions of Jarek Nabrzyski, Chris Sweet, Scott Szakonyi, Marc Gazda, Francis Rogg, Kendra Harding, Rob Duke, Kaitlyn Veenstra, and the Center for Research Computing at the University of Notre Dame.

## 9. References

[1] U.S. Energy Information Administration, (2014), Independent Statistics and Analysis. [Online] Available: http://www.eia.gov/tools/faqs/faq.cfm?id=86&t=1 (July 1, 2014)

[2] U. S. Department of Energy (2007/2008). Technical Report DOE/ EIA-0384 (2007). U. S. Department of Energy and Annual Energy Review.

[3] U. S. Department of Energy (2007/2008). Technical Report DOE/ EIA-0573(2007). Emissions of Greenhouse Gases in the United States.

[4] H.R. 6 (110th): Energy Independence and Security Act of 2007. [Online] Available: https://www.govtrack.us/congress/bills/110/hr6 (April 12, 2015)

[5] 2030 Inc., (2011), Architecture 2030. [Online] Available: http://architecture2030.org/ (July 1, 2014)

[6] Autodesk Inc., (2014), Autodesk: Revit®. [Online] Available: http://www.autodesk.com/products/autodesk-revit-family/overview/ (July 1, 2014)

[7] Autodesk Inc., (2014), Autodesk: Ecotect Analysis®. [Online] Available: http://usa.autodesk.com/ecotect-analysis/ (July 1, 2014)

[8] Athena Sustainable Materials Institute, (2014), Athena Impact Estimator®. [Online] Available: http://www.athenasmi.org/our-software-data/impact-estimator/ (July 1, 2014)

[9] Autodesk Inc., (2014), Autodesk: Green Building Studio®. [Online] Available: http://usa.autodesk.com/ecotect-analysis/ (July 1, 2014)

[10] U. S. Department of Energy (DOE-2), (2014). [Online] Available: http://energy.gov/ (July 1, 2014)

[11] U.S. Department of Commerce (BEES 4.0), (2014), National Institute of Standards and Technology: Engineering Laboratory. [Online] Available: http://www.nist.gov/el/economics/BEESSoftware.cfm (July 1, 2014)

[12] Timberlake, K., (2014), Tally LCA App for Autodesk Revit. [Online] Available: http://www.kierantimberlake.com/pages/view/95/tally/parent:4 (July 1, 2014)

[13] ISO 14040:2006 - Environmental management – Life cycle assessment – Principles and frame-work. [Online] Available: http:// www.iso.org/iso/catalogue_detail?csnumber=37456 (July 1, 2014)

[14] SuAT, (2014), Architecture and Sustainable Building Technologies. DPx-Design Performance Viewer, ETH. [Online] Available: http://www.suat.arch.ethz. ch/en/research/design-performance (July 1, 2014)

[15] GaBi Software, (2014), Product Sustainability Software®. PE International Inc. [Online] Available: www.gabi-software.com (July 1, 2014)

[16] Energy Star, (2014), United States Environmental Protection Agency (EPA). [Online] Available: http://www.energystar.gov (July 1, 2014)

[17] U.S. Green Building Council (USGBC), (2014), Leadership in Energy and Environmental Design (LEED). [Online] Available: http://www.usgbc.org/?CategoryID =19 (July 1, 2014)

[18] Lawrence Berkeley National Laboratory, (2013), EnergyPlus Version 8.2. [Online] Available: http://apps1.eere.energy.gov/buildings/tools_directory/software.cfm/ID=287/ (July 1, 2013)

[19] National Renewable Energy Laboratory (NREL), (2014), OpenStudio user documentation. 2012. [Online] Available: http://openstudio.nrel.gov/ (July 1, 2014)

[20] Hernandez, P., & Kenny, P. (2011). Development of a methodology for life cycle building energy ratings. Energy Policy, 39(6):3779–3788.

[21] Keoleian, G. A., Menery, D., & Curran, M., A. (1993). Life Cycle Design Guidance Manual. Office of Research and Development, Cincinnati, OH.

[22] Open Green Building XML Schema, (2014), gbXML: Open Green Building XML Schema: a Building Information Modeling Solution for Our Green World. [Online] Available: http://gbxml.org/ (July 1, 2014)

[23] Ishizaka, A. & Nemery, P. (2013). Multi-criteria Decision Analysis: Methods and Software. John Wiley & Sons, Ltd.

[24] Eames, M. E., Ramallo-Gonzalez, A. P., & Coley, D. A. (2013). Lumped parameter models for building thermal modelling: An analytical approach simplifying complex multi-layered constructions. Energy and Buildings, 60:174–184.

[25] Yu, N., Paolucci, S., Grenga, T., Salakij, S. (2014). Thermal Model of Green Scale Digital Design and Analysis Tool for Sustainable Buildings (BEAM). Aerospace and Mechanical Engineering Internal Report: University of Notre Dame.

[26] Kim, K., Kim, G., Yoo, D., Yu, J. (2012). Semantic material name matching system for building energy analysis. Automation in Construction, 30:242–255. DOI: 10.1016/j.autcon.2012.11.011.

[27] Kim, S. K., Moon, H. J., Choi, M. S., & Ryu, S. H. (2011). Case studies for the evaluation of interoperability between a BIM based architecture model and building performance analysis programs. Proceedings of Building Simulation 2011: 12th Conference of Intl Building Performance Simulation Association, Sydney, 14-16 November.

[28] Building System Laboratory. (1991). BLAST User Reference, (BSl).

[29] The Green Scale, (2014), Green Scale Research Project, University of Notre Dame. [Online] Available: www.greenscale.org (July 1, 2014)

[30] American Society of Heating, Refrigerating and Air Conditioning Engineers. (2007). ASHRAE Standard 90.1-2007. ASHRAE, Inc.

[31] International Finance Corporation, (2014), IFC: Green Buildings. [Online] Available: http://www.ifc.org/wps/wcm/connect/topics_ext_content/ifc_external_corporate_site/cb_home/sectors/green_buildings (July 1, 2014)

[32] National Institute of Building Sciences, (2014), National BIM Standard. [Online] Available: http://www.nationalbimstandard.org/ (July 1, 2014)

[33] Dixit, M. K., Fernandez-Solis, J., Lavy, S., & Culp, C. H. (2010). Identification of parameters for embodied energy measurement: A literature review. Energy & Buildings, 42(8):1238–1247.

[34] Vardeman, C., Krisnadhi, A. A., Cheatham, M., Janowicz, K., Ferguson, H., Hitzler, P., Buccellato, A., Thirunarayan, K., Berg-Cross, G., & Hahmann, T. (2014). An Ontology Design Pattern for Material Transformation. WOP2014 Patterns.

[35] W3C, (2014), Linked Data & Semantic Web. [Online] Available: http://www.w3.org/standards/semanticweb /data (September 1, 2014)

[36] Bertin, B., Scuturici, V., Pinon, J., Risler, E. (2012). Carbon DB: A semantic life cycle inventory database, 2683–2685.

[37] Kandil, A., Hastak, M., Bridges, P. S. D. (2014). An ontological approach to building information model exchanges in the precast/pre-stressed concrete industry, Volume 10 pp. 9780784412329–112.

[38] Bertin, B. Scuturici, M., Pinon, J. M., Risler, E. (2012). Semantic modelling of dependency relations between Life Cycle Analysis processes, 109–124.

[39] Ferguson, H., Vardeman, C., Buccellato, A. Capturing an Architectural Knowledge Base Utilizing Rules Engine Integration for Energy and Environmental Simulations. 2015 Proceedings of the Symposium on Simulation for Architecture and Urban Design. SimAUD 2015.

[40] Parmee, I. C., Abraham, J. A. Interactive Evolutionary Design. 2005 Knowledge Incorporation in Evolutionary Computation: Studies in Fuzziness and Soft Computing Volume 167 pp. 435-458.

**Appendix A**

**Green Scale Graphical User Interface (GUI) and Overall Application Architecture:**

The GST [Figure A.1] is a modular Python framework [Figure A.2], works to facilitate data interchange as a Revit plug-in, and is specifically designed to connect architectural models with various performance metrics. Modules are constructed for on-the-fly calculations of embodied energy and annual thermal performance ultimately achieving comparative analysis. The GST Interface [Figure A.1] is implemented as a Revit 2014 Plug-In and there is an underlying computational architecture presented in Figure A.2. The tool is "launched" by the user from the list of existing Plug-Ins provided in the Revit plugin user interface; the user is then required to choose surfaces from the architectural model of at least one construction type. For example, a construction type defines a set of geometries and/or drawing objects that define an assembly (i.e. the built combination of studs, insulation, and gypsum board). After the Plug-In launches, the GS user interface is presented from which the Python energy models are called [Figure A.1]. The tool is a Windows application that processes models by defining all possible construction types for a particular surface. It then generates a gbXML for each type found within the Revit API and presents a list of these choices to the user. The processing of the Python models occurs once for each assembly choice selected so as to create an iterative and extensible simulation tool [Figure A.1]. This programming sequence executes as a loop that can be repeated as desired by the user.

From the GS Interface layer, the Python based analysis models are launched separately allowing modularity as well as potential parallelism and extensibility; currently the module set includes life cycle inventory based Embodied Energy and Thermal Heat Flux metrics [Figure A.2]. Using each gbXML as provided by Revit, key-value dictionaries are constructed from the existing tags and values and accessed as needed for the calculations. A computational mapping of a given architectural designs parametric geometries is created at a level of granularity sufficient for any one of the analysis modules chosen to faithfully execute. Modules also require individual access to additional data sets and external entities; for example, the thermal model needs access to Energy Plus climatology data. The capability to create log files for each of the modules for debugging and additional calculation information is also implemented.

The tool uses a SQLite3[3] database called the "GS Material Property Database" that is queried from within the running Python models via a Django[4] based RESTful web service running on a remote Linux server. The RESTful interface returns JSON[5] data serialization that is added to the local in-memory dictionaries. To limit the size of web transfers such that interruptions from web connectivity are less likely, JSON processes requests for data in sets of thirty materials for as many materials exist in the tree.

Once each one of the Python modules has completed execution for any assembly versions selected to undergo comparative analysis, the aggregated calculation results are transferred back to the user via standard output. At this point, the next gbXML would be sent for Python module processing and also the updated results of the currently finished model are displayed in one of the GS interface windows [Figure A.1]. All of the results are reported in the form of graphics such as text tables, charts, graphs, and PDFs to name a few [Figure A.1].

As a result, the GST exists within a modular framework that facilitates multi-metric calculations and simultaneous comparative analysis. As modules are added to the tool and the material database expands, the application will be able to connect to other data sources that conform to a RESTful API - collecting new material data on-the-fly as needed by the models. Data aggregations can even be extended to use cloud-based or crowd-sourced datasets as the community grows [Figure A.2]. These are only a few of the long-term benefits that exist due to the inherent nature of rule-based processing of properly structured data.

---

[3] SQLite3: http://sqlite.org
[4] Django: https://www.djangoproject.com
[5] JSON: http://www.json.org

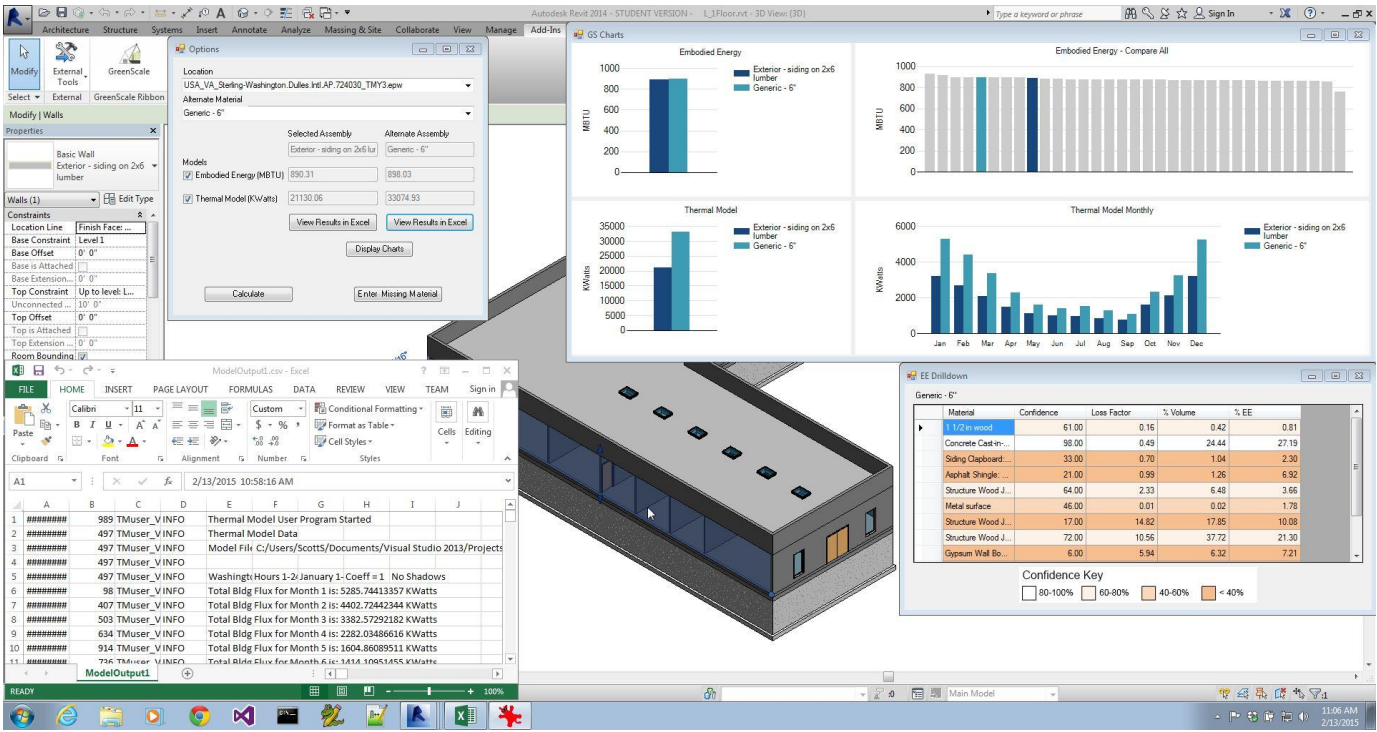

**Figure A.1: Green Scale Tool Interface from Revit Plug-In**

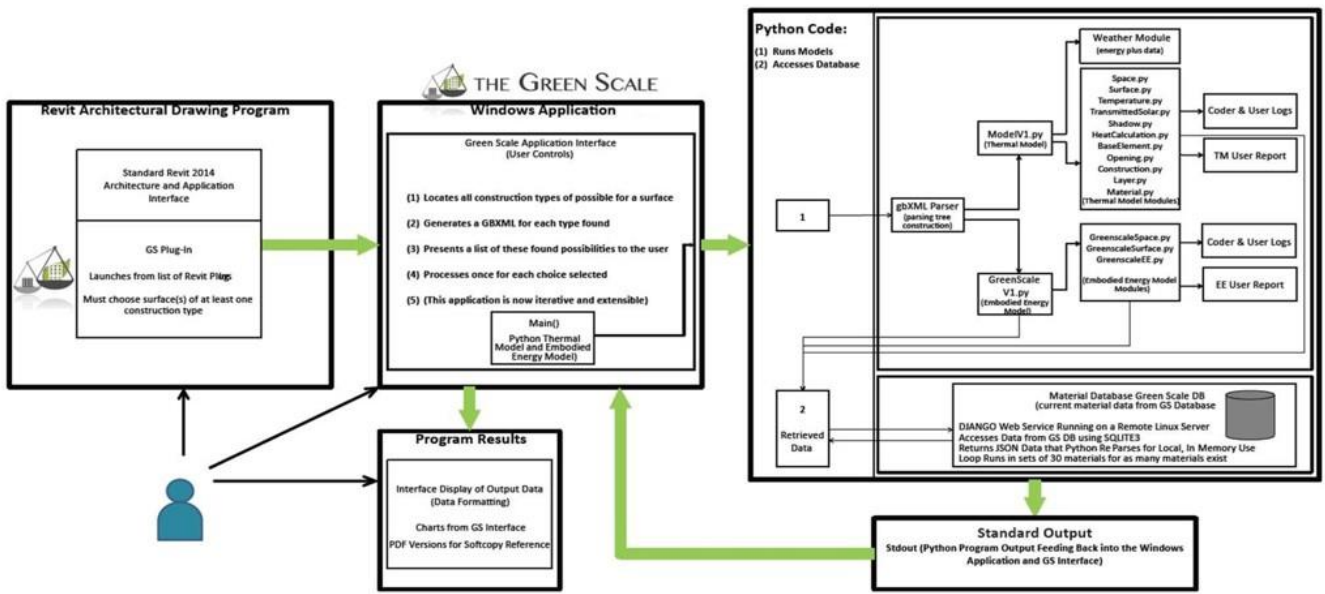

**Figure A.2: Green Scale Tool Architecture**

# Appendix B

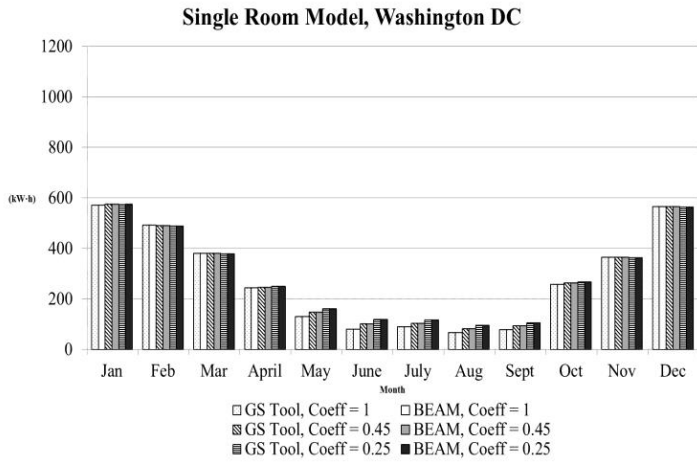

**Figure B.1**

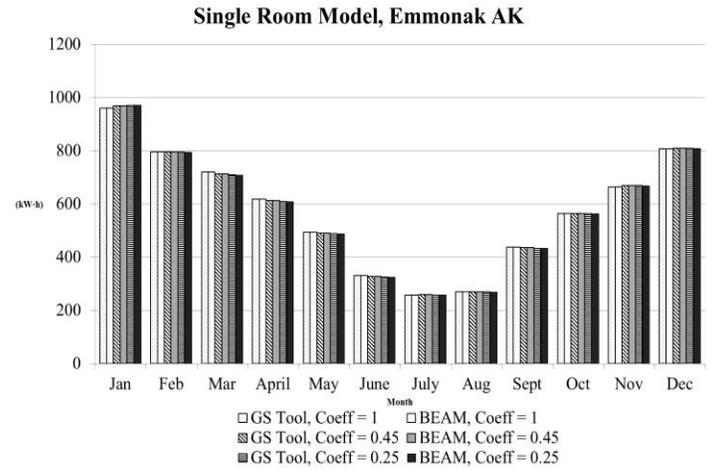

**Figure B.3**

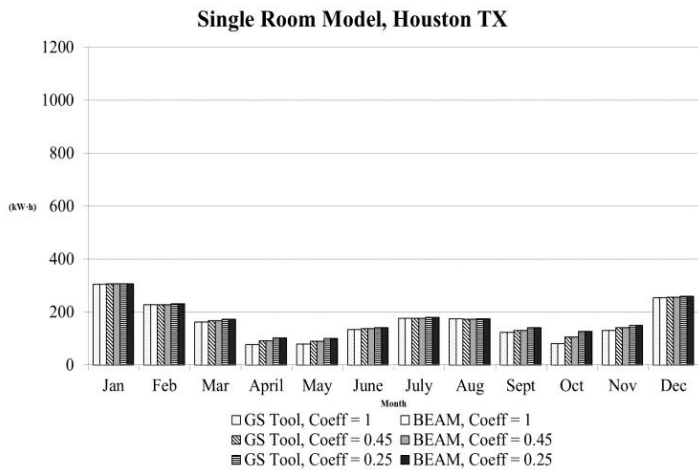

**Figure B.2**

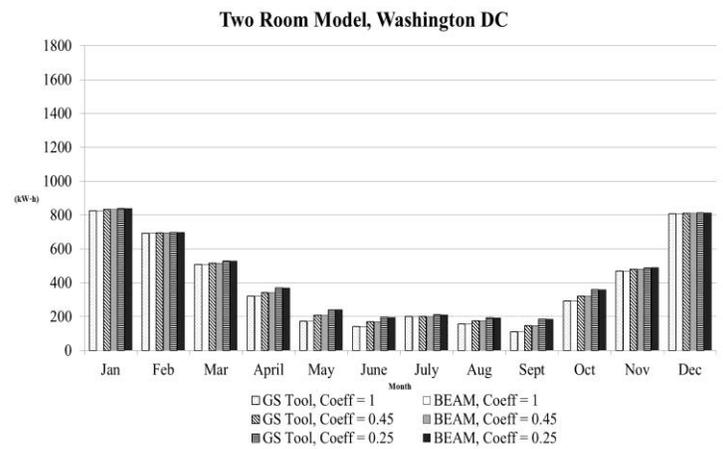

**Figure B.4**

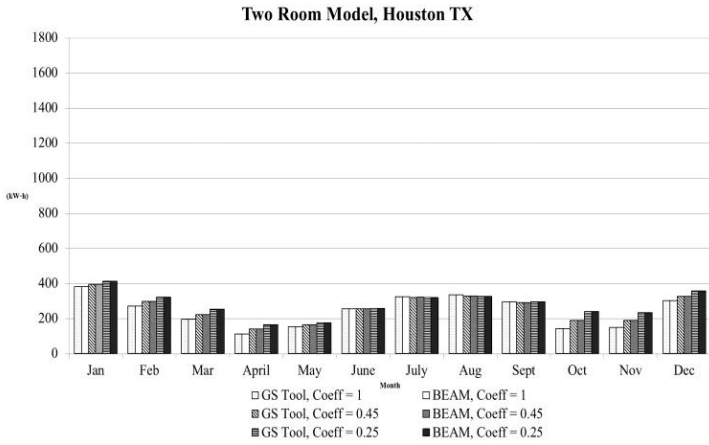

**Figure B.5**

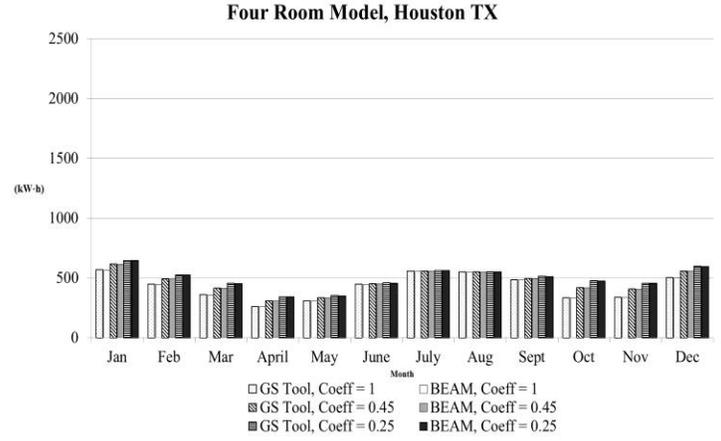

**Figure B.8**

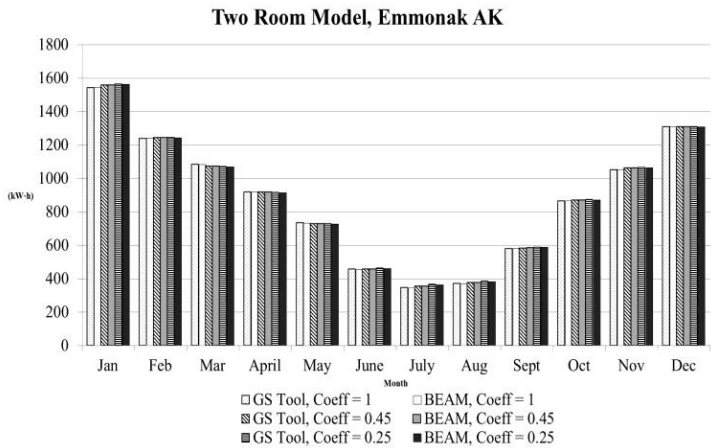

**Figure B.6**

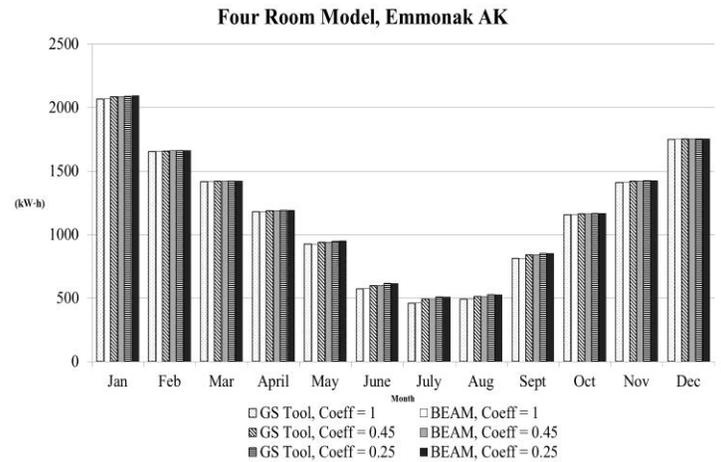

**Figure B.9**

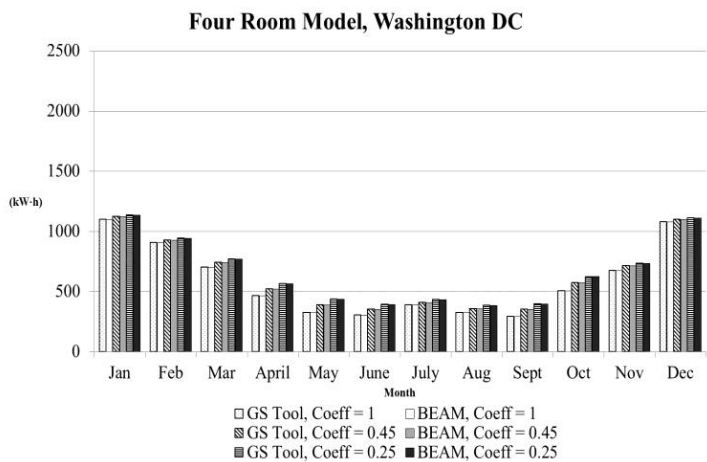

**Figure B.7**